\documentclass[aps,prd,reprint,twocolumn,superscriptaddress,showpacs]{revtex4-1}
\usepackage{graphicx}
\usepackage{mathrsfs}
\usepackage{bm}
\usepackage{amsmath}
\usepackage{dcolumn}
\usepackage{epstopdf}
\usepackage{dsfont}
\usepackage{amssymb}
\usepackage{tabularx}
\usepackage{array}
\usepackage{float}
\usepackage{color}
\usepackage{epstopdf}
\usepackage{mathrsfs}
\usepackage[colorlinks, linkcolor=blue,anchorcolor=blue,citecolor=blue,urlcolor=blue]{hyperref}

\begin{document}
\title{Almost ideal nodal-loop semimetal in monoclinic CuTeO$_3$ material}

\author{Si Li}
\email{S. Li and Y. Liu contributed equally to this work.}
\affiliation{Beijing Key Laboratory of Nanophotonics and Ultrafine Optoelectronic Systems, School of Physics,
Beijing Institute of Technology, Beijing 100081, China}
\affiliation{Research Laboratory for Quantum Materials, Singapore University of Technology and Design, Singapore 487372, Singapore}

\author{Ying Liu}
\email{S. Li and Y. Liu contributed equally to this work.}
\affiliation{Research Laboratory for Quantum Materials, Singapore University of Technology and Design, Singapore 487372, Singapore}

\author{Botao Fu}
\affiliation{Beijing Key Laboratory of Nanophotonics and Ultrafine Optoelectronic Systems, School of Physics,
Beijing Institute of Technology, Beijing 100081, China}

\author{Zhi-Ming Yu}
\affiliation{Research Laboratory for Quantum Materials, Singapore University of Technology and Design, Singapore 487372, Singapore}

\author{Shengyuan A. Yang}
\affiliation{Research Laboratory for Quantum Materials, Singapore University of Technology and Design, Singapore 487372, Singapore}

\author{Yugui Yao}\email{ygyao@bit.edu.cn}
\affiliation{Beijing Key Laboratory of Nanophotonics and Ultrafine Optoelectronic Systems, School of Physics,
Beijing Institute of Technology, Beijing 100081, China}

\begin{abstract}
Nodal-loop semimetals are materials in which the conduction and valence bands cross on a one-dimensional loop in the reciprocal space. For the nodal-loop character to manifest in physical properties, it is desired that the loop is close to the Fermi level, relatively flat in energy, simple in its shape, and not coexisting with other extraneous bands. Here, based on the first-principles calculations, we show that the monoclinic CuTeO$_3$ is a realistic nodal-loop semimetal that satisfies all these requirements. The material features only a single nodal loop around the Fermi level, protected by either of the two independent symmetries: the $\mathcal{PT}$ symmetry and the glide mirror symmetry. The size of the loop can be effectively tuned by strain, and the loop can even be annihilated under stain, making a topological phase transition to a trivial insulator phase.
Including the spin-orbit coupling opens a tiny gap at the loop, and the system becomes a $\mathbb{Z}_2$ topological semimetal with a nontrivial bulk $\mathbb{Z}_2$ invariant but no global bandgap. The corresponding topological surface states have been identified. We also construct a low-energy effective model to describe the nodal loop and the effect of spin-orbit coupling.
\end{abstract}

\maketitle
\section{Introduction}
The study of topological states of matter have been attracting significant interest in the current condensed matter physics research. Topological insulators which possess insulating bulk states and robust metallic surface states have been extensively studied~\cite{hasan2010,qi2011}. Recently, topological metals and semimetals have emerged as a new research focus~\cite{Chiu2016,Burkov2016,armitage2018}. Their band structures exhibit nontrivial band crossings near the Fermi energy, around which the low-energy quasiparticles behave differently from the conventional Schr\"{o}dinger-type fermions. For instance, Weyl and Dirac semimetals host isolated twofold and fourfold degenerate linear band-crossing points, respectively, where the electronic excitations are analogous to the relativistic Weyl and Dirac fermions~\cite{Volovik2003,Wan2011,Murakami2007,Burkov2011,Young2012,Wang2012b,Wang2013b,Zhao2013c,Yang2014a,Weng2015,Liu2014c,Borisenko2014,Lv2015,Xu2015a}, making it possible to simulate interesting high-energy physics phenomena in condensed matter systems~\cite{Nielsen1983,son2013,guan2017a}.

The conduction and valence bands may also cross along a one-dimensional (1D) loop in the Brillouin zone (BZ). Such nodal loops can be protected by symmetries and may lead to drumhead-like surface bands. Several interesting properties have been predicted for nodal-loop semimetals, such as the anisotropic electron transport~\cite{Mullen2015}, the unusual optical response and circular dichorism~\cite{Carbotte2016,Ahn2017,Liu2018}, the possible surface magnetism and superconductivity~\cite{Heikkilae2011,Li2016,Chan2016,Wang2017b,Liu2017}, the anomalous Landau level spectrum~\cite{Rhim2015,Lim2017}, and density fluctuation plasmons and Friedel oscillations~\cite{Rhim2014}. Quite a few realistic materials have been proposed as nodal-loop semimetals~\cite{Yang2014,Weng2015c,Mullen2015,Yu2015,Kim2015a,Chen2015,Chen2015a,Fang2016,Chan2016,Bian2016,Schoop2016,Gan,Li2017,Zhang2017,Huang2017,Jiao2017,Li2018},
and interestingly, Dirac and Weyl loops have also been proposed in cold-atom optical lattices~\cite{xu2016}. However, to unambiguously identify the predicted features due to the nodal loops in these systems still remains challenging, partly because the low-energy band structures of the realistic materials suffer from various drawbacks. As a good nodal-loop semimetal, the material should at least satisfy the following requirements. First, the nodal loop should be close to the Fermi level. Second, the energy variation along the loop should be as small as possible. Third, the loop has a relatively simple shape, and it is better that only a single loop appears at low-energy. Fourth, it is crucial that no other extraneous bands are nearby in energy, since otherwise they may complicate the interpretation of measured properties (like for the transport coefficients) ~\cite{gao2017intrinsic}. Hence, to facilitate the experimental exploration of the nodal-loop semimetals, an urgent task is to identify realistic materials that satisfy these requirements.

In this work, based on first-principles calculations and symmetry analysis, we predict that the monoclinic CuTeO$_3$ is a nodal-loop semimetal which satisfies all the above-mentioned requirements. The material features a single nodal loop close to the Fermi level in its low-energy band structure. The loop is almost flat in energy and there is no other extraneous band crossing the Fermi level. The stability of the loop is protected by either of the two independent symmetries in the absence of spin-orbit coupling (SOC): the $\mathcal{PT}$ symmetry and the glide mirror symmetry of the system. We show that lattice strain can effectively tune the shape of the loop, and even annihilate the loop to make a topological phase transition into a trivial insulator phase. SOC opens a gap at the nodal loop, and transforms the system into a  $\mathbb{Z}_2$ topological semimetal which has a nontrivial bulk  $\mathbb{Z}_2$ invariant but no global bandgap. The corresponding nontrivial surface states are revealed. We also construct a low-energy effective model to capture the nodal-loop as well as the effect of SOC. Since the SOC is negligible for this material, the nodal-loop features should clearly manifest in experimental measurements. Our result suggests an almost ideal platform for experimentally exploring the intriguing properties of nodal-loop semimetals.

\section{CRYSTAL STRUCTURE}

CuTeO$_3$ belongs to the tellurium(IV)-oxygen compounds. It may crystalize into two typical structures: the orthorhombic structure and the monoclinic structure, which are denoted as structure I and II~\cite{pertlik1987}, respectively. In this work, we focus on the monoclinic CuTeO$_3$, which has been successfully synthesized by a hydrothermal method and shown to be stable at ambient condition~\cite{pertlik1987}. The chemical formula and the crystal structure have been determined by a quantitative analysis of the X-ray diffraction data (the crystal structure data are $a = 5.965$ \AA, $b =5.214$ \AA, $c = 9.108$ \AA, and $\gamma=95.06^\circ$, where $\gamma$ is the angle between $a$ and $b$ axis) ~\cite{pertlik1987}.

\begin{figure}[t!]
\includegraphics[width=8.8cm]{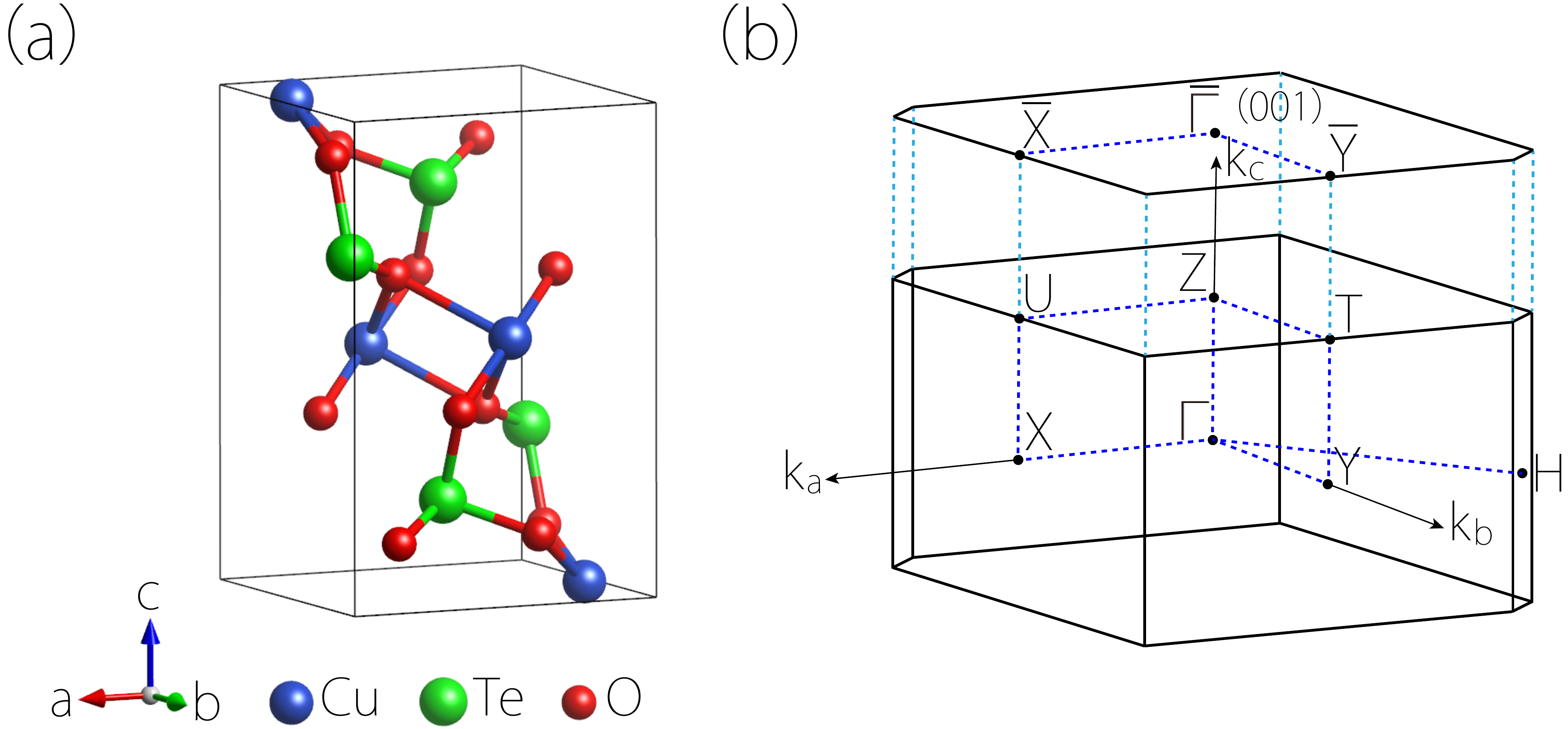}
\caption{(a) Crystal structure of the monoclinic CuTeO$_3$. The figure shows the unit cell of the structure. (b) The bulk Brillouin zone and the projected surface Brillouin zone of the (001) plane. The high-symmetry points are labeled.}
\label{fig1}
\end{figure}

The monoclinic CuTeO$_3$ has a structure with space group No.~14 ($P2_1/c$) [see Fig.~\ref{fig1}(a)], which can be generated by the following symmetry elements: the inversion $\mathcal{P}$ [inversion center located at $(a/2,b/2,c/2)$] and the glide mirror $\widetilde{\mathcal{M}}_{z}:\ (x,y,z)\rightarrow (x+\frac{1}{2},y+\frac{1}{2},-z+\frac{1}{2})$. Here the tilde denotes a nonsymmorphic operation, which involves a translation with fractional lattice parameters. In addition, the material has been found to show no magnetic ordering~\cite{pertlik1987}, so the time reversal symmetry $\mathcal{T}$ is also preserved. The crystal structure of monoclinic CuTeO$_3$ and the BZ are schematically shown in Fig.~\ref{fig1}.

\section{First-principles Methods}

We performed first-principles calculations based on the density functional theory (DFT) using the projector augmented wave method as implemented in the Vienna ab initio simulation package~\cite{Kresse1994,Kresse1996,PAW}. The exchange-correlation functional was modeled within the generalized gradient approximation (GGA) with the Perdew-Burke-Ernzerhof (PBE) realization~\cite{PBE}. The cutoff energy was set as 550 eV, and the BZ was sampled with a $\Gamma$-centered $k$ mesh of size $20\times 20\times 12$. The experimental lattice parameters ($a = 5.965$ \AA, $b =5.214$ \AA, $c = 9.108$ \AA )~\cite{pertlik1987} were adopted in the calculation. The band structure result was further checked by using the more accurate approach with the modified Becke-Johnson (mBJ) potential~\cite{tran2009}. We confirm that the essential features including the nodal loop remain the same as in the GGA result (see Appendix A). As Cu($3d$) orbitals may have correlation effects, we also tested our result by using the GGA$+U$ method~\cite{anisimov1991}, and the results are discussed in Section V. In order to study the topological surface states, localized Wannier functions were constructed by projecting the Bloch states onto atomic-like trial orbitals without an iterative procedure~\cite{Marzari1997,Souza2001,Huang2017}. Using the Wannier functions, the surface spectra were calculated via the iterative Green's function method~\cite{Green} as implemented in the WannierTools package~\cite{Wu2017}.

\section{RESULTS}

\subsection{Nodal-loop semimetal in CuTeO$_3$}

\begin{figure}[b!]
\includegraphics[width=8.8cm]{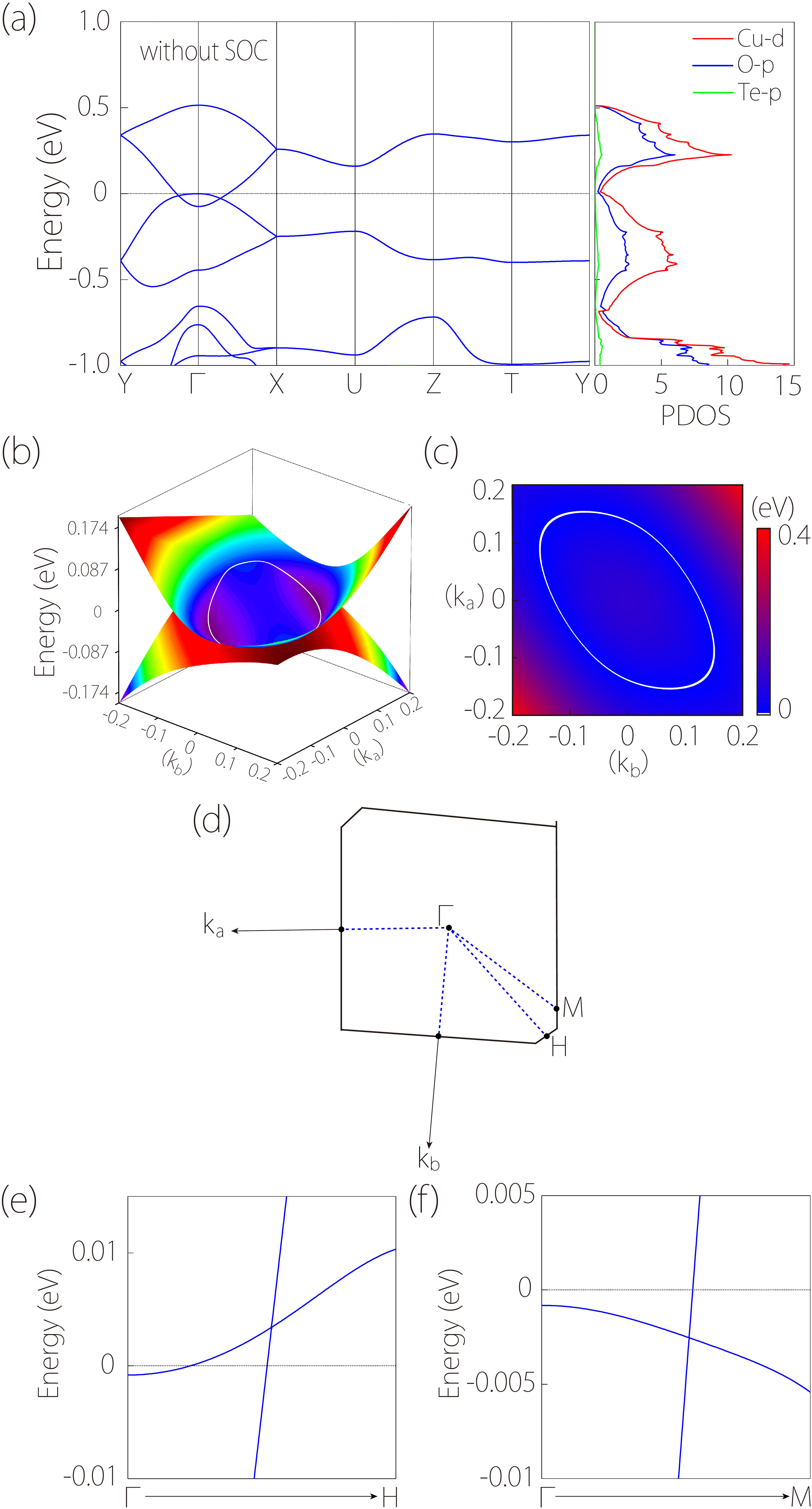}
\caption{(a) Band structure for the monoclinic CuTeO$_3$. The right panel shows the projected density of states (PDOS). The result here is in the absence of SOC. (b) Band dispersion in the $k_z=0$ plane near the $\Gamma$ point. The crossing between the two low-energy bands forms a nodal loop. (c) Shape of the nodal loop (white curve)
obtained from the DFT calculation. The color map indicates the local gap between the two crossing bands. (d) Illustration of a few paths in the $k_z=0$ plane. (e) and (f) are the zoom-in images for the low-energy bands along two paths indicated in (d), showing that the points in a small section on the nodal loop (e.g. along $\Gamma$-H) is type-II.}
\label{fig2}
\end{figure}

We first consider the electronic band structure of monoclinic CuTeO$_3$ (hereafter referred to as simply CuTeO$_3$) in the absence of SOC. The band structure from DFT calculation is shown in Fig.~\ref{fig2}(a), along with the projected density of states (PDOS). From PDOS, one clearly observes that the system is a semimetal: it has zero bandgap but the density of states at the Fermi level is very small. The low-energy states near the Fermi level are mainly from the Cu($3d$) orbitals and the O($2p$) orbitals.

In the band structure, there appear linear band-crossing points along the Y-$\Gamma$ and $\Gamma$-X paths very close to the Fermi level [see Fig.~\ref{fig2}(a)]. Actually, these points are not isolated. Via a careful scan of the crossing points in the BZ, we find that the two points belong to a nodal loop centered around the $\Gamma$ point in the $k_z=0$ plane, formed by the crossing between the conduction and the valence bands.
This is more clearly shown in Fig.~\ref{fig2}(b), where we plot the dispersion of the two bands in the $k_z=0$ plane around $\Gamma$. One directly observes that they cross along a 1D loop in this plane. The shape of the nodal loop obtained from the scan is shown in Fig.~\ref{fig2}(c). These results demonstrate that the material is a nodal-loop semimetal with a single nodal loop close to the Fermi level in the band structure. The loop is quite flat in energy, with an energy variation less than 0.04 eV. More importantly, the band structure is clean in the sense that there is no other extraneous band near the Fermi level. Thus, CuTeO$_3$ satisfies the conditions we listed at the Introduction section for a good nodal-loop semimetal.

Depending on the slope of the two crossing bands, a nodal point can be classified as type-I or type-II~\cite{Xu2015,Soluyanov2015}. Recently, Li \emph{et al.} proposed the concepts of type-II and hybrid nodal loops~\cite{Li2017}, based on the type of the nodal points that make up the loop. A loop is type-I (type-II) if all the points on the loop are type-I (type-II) in the 2D transverse dimensions. If a loop contains both type-I and type-II points, then it corresponds to a hyrbid type. Several interesting physical properties have been predicted for each type of the loops~\cite{heikkila2015,hyart2016,gao2017,zhang2018}. Here, we also determine the type of the nodal loop found in CuTeO$_3$. By scanning the dispersion around the loop, we find that although most points on the loop are of type-I [Fig.~\ref{fig2}(f)], there are small sections of the loop where the points are of type-II [Fig.~\ref{fig2}(e)]. Thus, the loop contains both type-I and type-II points, and according to the definition, it belongs to the hybrid type.

\begin{figure}[t!]
\includegraphics[width=8.8cm]{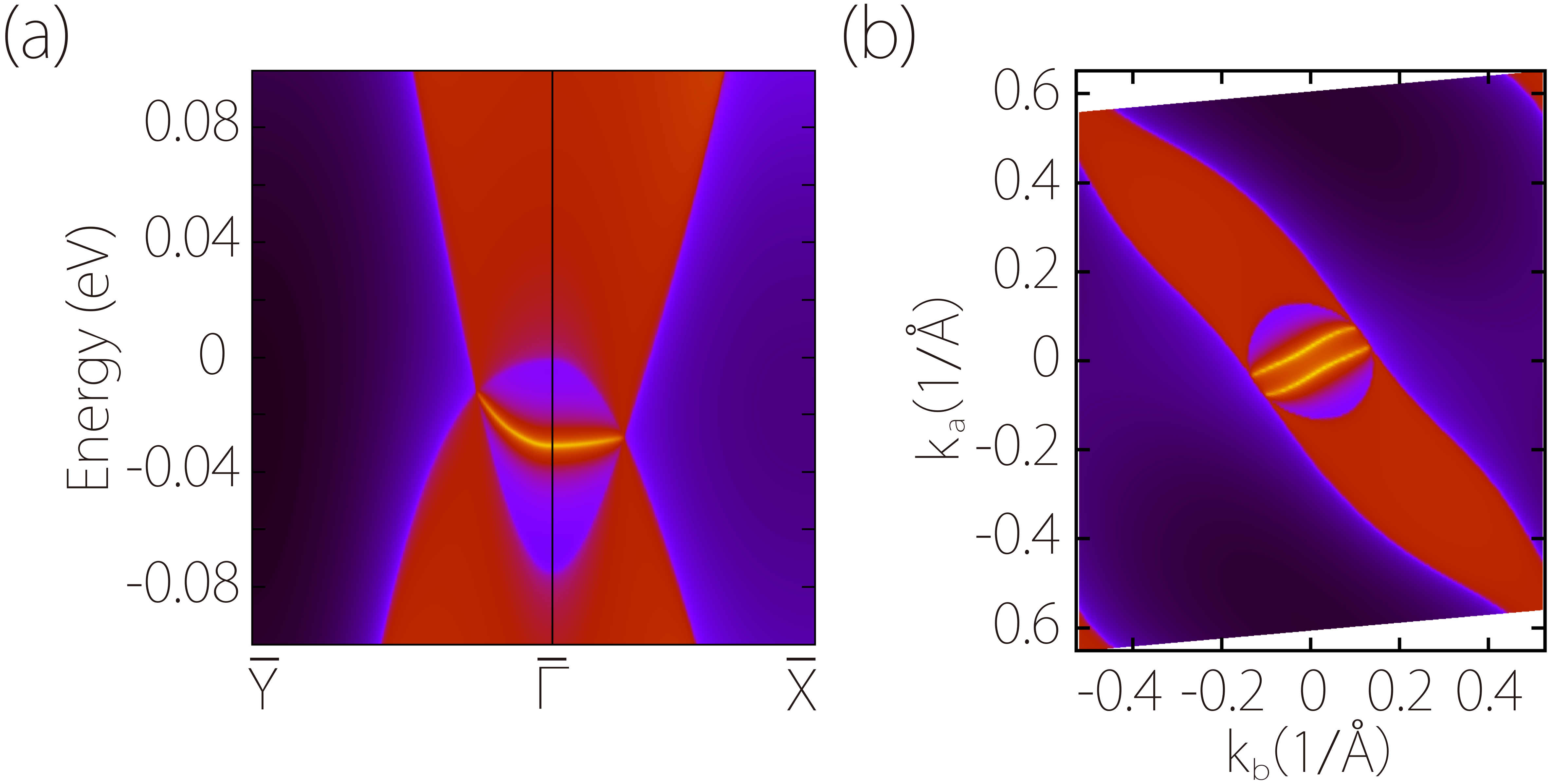}
\caption{(a) Projected spectrum on the (001) surface, and (b) the corresponding constant energy slice at $-0.03$ eV.}
\label{fig3}
\end{figure}

Nodal loops usually display drumhead-like surface states at the sample surface, on which the loop has a nonzero projected area~\cite{Yang2014,Weng2015c}. In Fig.~\ref{fig3}, we show the spectrum for the (001) surface. One indeed observes a surface band connecting the nodal points that correspond to the loop in the bulk [see Fig.~\ref{fig3}(a)]. And the surface states lie within the surface projected nodal loop, as seen in Fig.~\ref{fig3}(b) for a constant energy slice at $-0.03$ eV. It is worth noting that in Fig.~\ref{fig3}(b), the surface states in the constant energy slice appear as open arcs, which seems peculiar as one typically expects a closed curve, as found in most previous works. However, such result is not unreasonable, because the shape of the surface states in the constant energy slice depends on the detailed dispersion of the surface band. Consider a surface band having a saddle-like shape in its dispersion, then on a constant energy slice, the surface states naturally appear as sections of hyperbolic curves (open arcs).

\subsection{Symmetry protection}

The nodal loop in CuTeO$_3$ is protected by either of the two independent symmetries: the $\mathcal{PT}$ symmetry and glide mirror symmetry $\widetilde{\mathcal{M}}_{z}$, when SOC is absent.

 In the absence of SOC, spin is a dummy degree of freedom, and the usual treatment is to simply drop the spin labels and the trivial spin degeneracy. Hence, for the current system, without SOC, the nodal loop can be regarded as formed by the crossing between two (spinless) bands. Then the presence of $\mathcal{PT}$ symmetry requires that the Berry phase for any closed 1D manifold to be quantized into multiples of $\pi$~\cite{zhao2016}. Here, the Berry phase
\begin{equation}
\gamma_\ell=\sum_{n\in \text{occ.}}\oint_\ell \langle u_n(\bm k)|i\nabla_{\bm k}u_n(\bm k)\rangle d\bm k
\end{equation}
is defined for a locally gapped spectrum along a closed path $\ell$, $|u_n(\bm k)\rangle$ is the periodic part of the Bloch eigenstate, and the band index $n$ is summed over the occupied valence bands below the local gap. Hence, $\gamma_\ell/\pi$ mod $2$ defines a 1D $\mathbb{Z}_2$ invariant. For a nodal loop, $\gamma_\ell$ for a closed path $\ell$ encircling the loop is $\pm \pi$, hence protecting the loop against weak perturbations from opening a gap. In DFT calculations, we have also numerically checked that such Berry phase is nontrivial.

Another protection is from the glide mirror symmetry $\widetilde{\mathcal{M}}_{z}$. The loop lies in the $k_z=0$ plane, which is invariant under $\widetilde{\mathcal{M}}_{z}$. Hence, each state in this plane is also an eigenstate of $\widetilde{\mathcal{M}}_{z}$, with a well-defined eigenvalue $g_z=\pm e^{-ik_x/2-ik_y/2}$. The nodal loop is protected if the two crossing bands have opposite $g_z$ in the $k_z=0$ plane, which is indeed the case as verified by our DFT calculations (the conduction band has positive eigenvalues, and the valence band has negative eigenvalues). The presence of $\widetilde{\mathcal{M}}_{z}$ symmetry further pins the nodal loop to be within the $k_z=0$ plane.

\begin{figure}[t]
\includegraphics[width=8.6cm]{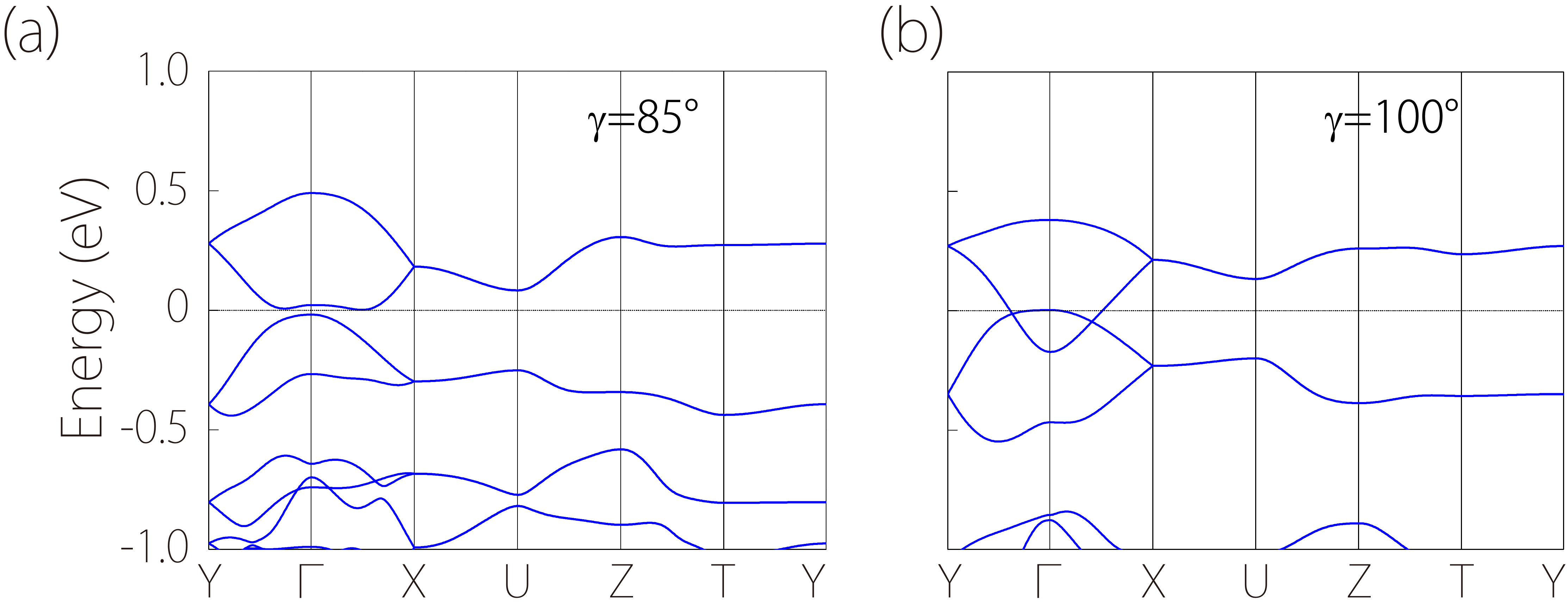}
\caption{Evolution of the loops when changing the angle $\gamma$ between $a$ and $b$ axis, for (a) $\gamma=85^\circ$ and (b) $\gamma=100^\circ$.}
\label{fig4}
\end{figure}

\begin{figure}[b!]
\includegraphics[width=8.6cm]{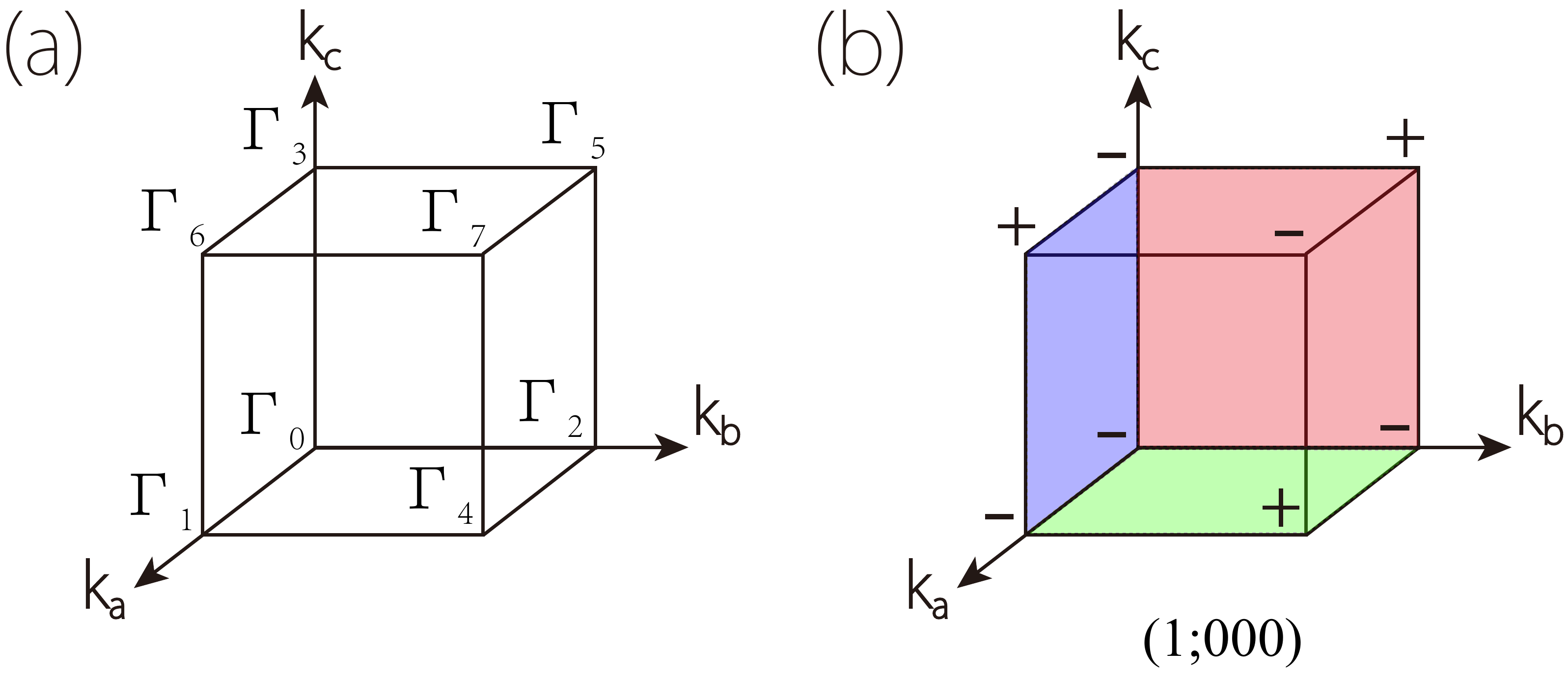}
\caption{Analysis of the parity eigenvalues. The eight TRIM points $\Gamma_i$ ($i=0,1,\cdots,7$) are schematically labeled in (a). (b) shows the corresponding parity eigenvalues $\xi_i$. The shaded regions indicate the three nontrivial invariant planes where the nodal loop pierces through. }
\label{fig5}
\end{figure}

The two symmetries are independent, so the loop is protected as long as one of them is preserved. We have checked this point by making lattice distortions that break one while maintaining the other. This also offers possibility to tune the nodal loop by strain engineering. The location and the shape of the loop can be effectively tuned by strain, if the applied strain preserves one of the two symmetries. For example, let's consider a lattice deformation by varying the angle $\gamma$ between the $a$ and $b$ axis, corresponding to a type of shear strain. As shown in Fig.~\ref{fig4}, one observes that the shape of the loop varies with strain, and it becomes larger when the angle is increased. Importantly, at decreased angle, the loop shrinks as the band inversion at $\Gamma$ is reduced. For $\gamma=85^\circ$, the loop is completely removed, and the system has made a topological phase transition into a trivial insulator phase.

 In addition, since the system preserves inversion symmetry, we can also analyze the topology of the nodal loop using the method discussed in Ref.~\cite{Kim2015a}. In this method, we calculate the product of parity eigenvalues $\xi_i$ for the occupied states at the eight time reversal invariant momenta (TRIM) $\Gamma_i$, where $i=0,1,\cdots, 7$. The results are indicated in Fig.~\ref{fig5}. Then we can obtain the $\mathbb{Z}_2$ invariant $\omega(C_{abcd})=\xi_a\xi_b\xi_c\xi_d$ for the six invariant $S_{abcd}$ planes. Three of the six planes have nontrivial $\omega(C)=-1$, as marked by the shaded regions in Fig.~\ref{fig5}(b). This is consistent with the fact that the loop pierces through these three planes.  This analysis also gives the $\mathbb{Z}_2$ invariant $(\nu_0;\nu_1\nu_2\nu_3)$ for a topological insulator once SOC is included, as we discuss in a while.

\subsection{Effect of SOC}

\begin{figure}[t!]
\includegraphics[width=8cm]{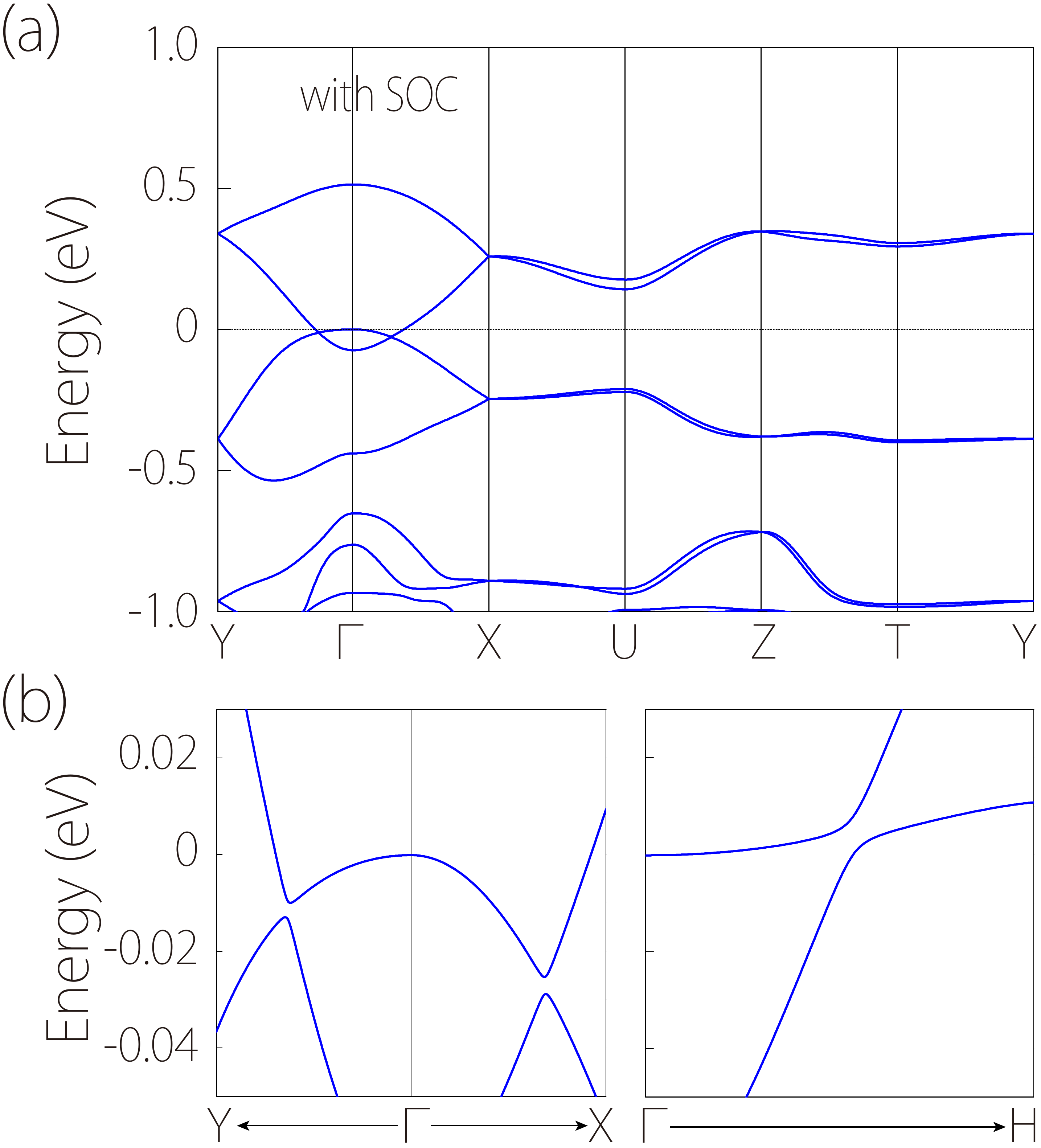}
\caption{(a) Band structure of the monoclinic CuTeO$_3$ with SOC included. (b) Zoom-in band structures around several paths, showing that a small gap is opened at the original nodal loop by SOC.}
\label{fig6}
\end{figure}

\begin{figure}[b!]
\includegraphics[width=5.5cm]{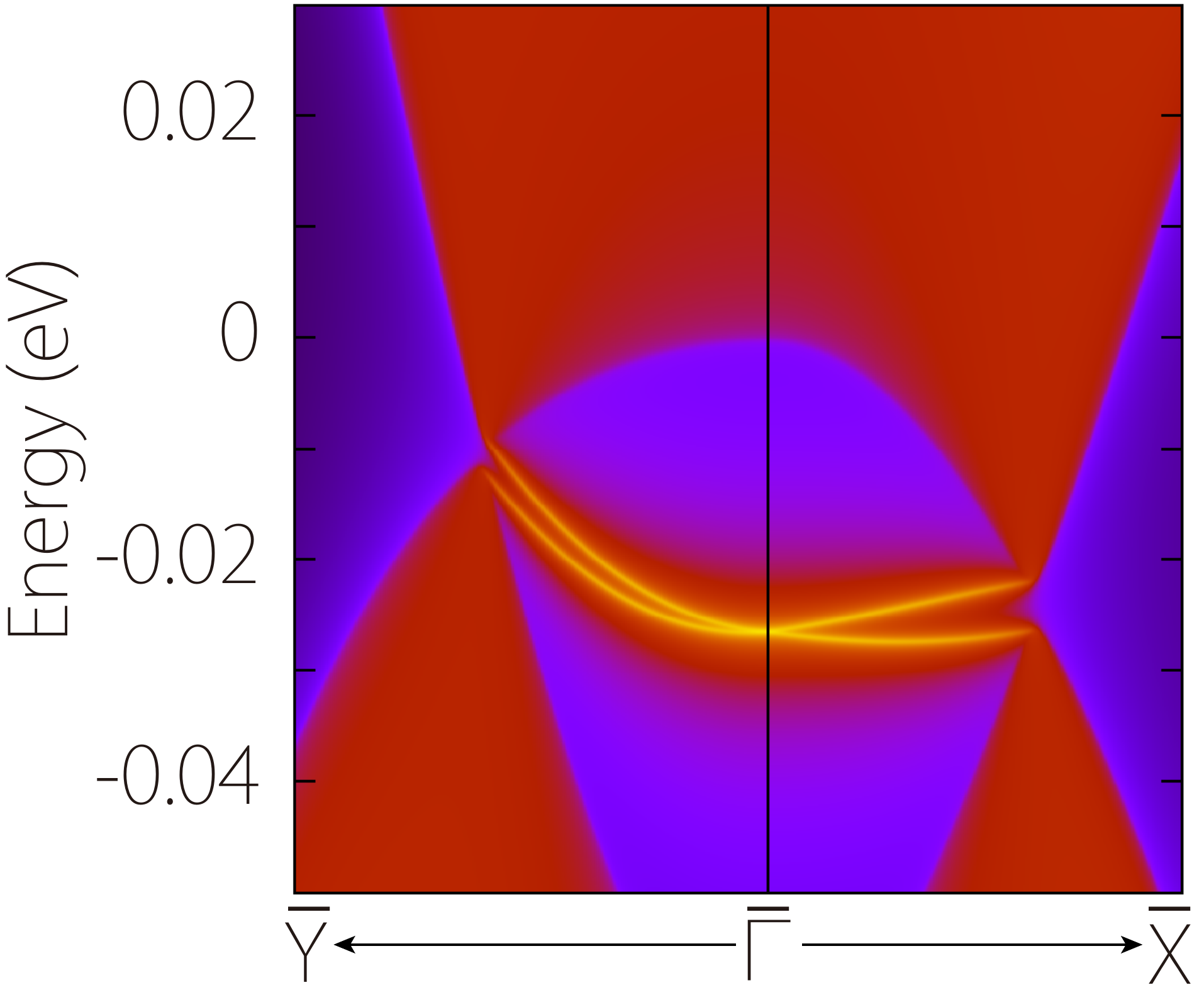}
\caption{ Topological surface states on the (001) surface when SOC is included.}
\label{fig7}
\end{figure}

Next, we turn to the band structure with SOC. Because the low-energy states are mainly from the Cu and O atomic orbitals, the SOC effect is expected to be small since these are light elements. This is verified in the DFT calculation. The band structure with SOC is plotted in Fig.~\ref{fig6}(a). One observes that it is quite similar to the result without SOC in Fig.~\ref{fig2}(a). Here, each band is at least doubly degenerate due to the $\mathcal{PT}$ symmetry. This is because each $k$ point is invariant under $\mathcal{PT}$, and with SOC, we have $(\mathcal{PT})^2=-1$, leading to a Kramers-like degeneracy for each band at each $k$ point. It should be noted that even if the individual $\mathcal{P}$ and $\mathcal{T}$ are broken, as along as the combined symmetry $\mathcal{PT}$ holds, the double degeneracy will still be there (such as in some magnetic materials). Zooming in the crossings corresponding to the nodal loop [see Fig.~\ref{fig6}(b)], one can see that there is a very small gap (about 0.0075 eV) opened at the original nodal loop.

After including the SOC, there is a local gap between conduction and valence bands at every $k$ point. Hence, we can have a well-defined $\mathbb{Z}_2$ invariant for the valence bands~\cite{pan2014,Guan2017}, just like that for the 3D topological insulators. Since band inversion occurs at the $\Gamma$ point, one naturally expects that this $\mathbb{Z}_2$ invariant is nontrivial. {The material here has inversion symmetry, so the  $\mathbb{Z}_2$ invariant can be conveniently evaluated by analyzing the parity eigenvalues at the TRIM points~\cite{fu2007}. This in fact has been done in the previous section (see Fig.~\ref{fig5}), and we find that $\mathbb{Z}_2=(1;000)$, which is indeed nontrivial.} Note that there is no global bandgap for the spectrum---the bandgap is closed indirectly, so in a strict sense the system is a $\mathbb{Z}_2$ topological semimetal~\cite{pan2014,Guan2017}. Like topological insulators, such state also possesses spin-momentum-locked topological surface states. In Fig.~\ref{fig7}, we show the surface spectrum for the (001) surface. Compared with the result in Fig.~\ref{fig3}(a), one observes that the drumhead surface states are split by the SOC and evolve into the spin-polarized surface states for the $\mathbb{Z}_2$ topological semimetal.

Here, it should be mentioned that the SOC effect is quite small for CuTeO$_3$. At energy scales larger than the SOC gap ($\sim 0.0075$ eV), the SOC effect has negligible influence on measured physical properties, and the material may be well described by a nodal-loop semimetal.

\subsection{Low-energy effective model}

\begin{figure}[b]
\includegraphics[width=8.6cm]{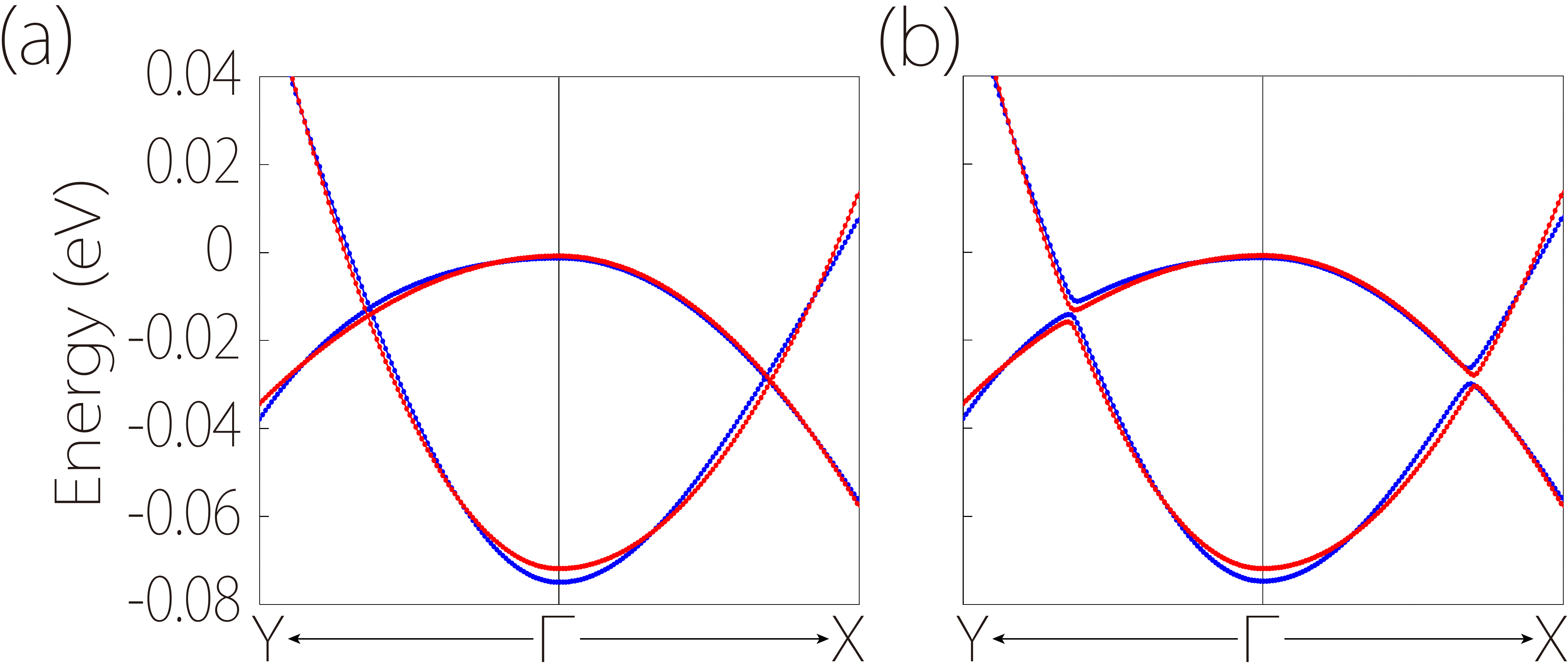}
\caption{Dispersions around the $\Gamma$ point fitted by the effective model. (a) is for the band structure without SOC, and (b) is for the band structure with SOC included.  The blue dots are results from DFT calculations, and the red dots are the fitting by the $k\cdot p$ model in Eq.~(\ref{kp1}) and Eq.~(\ref{kp2}).}
\label{fig8}
\end{figure}

To further characterize the nodal-loop semimetal phase and to understand the effect of SOC, we construct a low-energy effective ($k\cdot p$) model based on the symmetry requirements.

We first consider the case without SOC. The two low-energy states at the $\Gamma$ point belong to the $\Gamma_2^{+}$ and $\Gamma_2^{-}$ representations of the $C_{2h}$ point group at $\Gamma$. Using them as basis, we can obtain a two-band model that is constrained by symmetry. To capture the nodal loop, we expand the model at $\Gamma$ to leading order terms in each wave-vector component $k_i$, which gives
\begin{equation}\label{kp1}
\begin{split}
\mathcal{H}_0(\boldsymbol{k})=M(\boldsymbol{k})\tau_0+A(\boldsymbol{k})\tau_z+B(\boldsymbol{k})\tau_y,
\end{split}
\end{equation}
where the Pauli matrices $\tau_i$ are in the space spanned by the two basis states at $\Gamma$,
$M(\boldsymbol{k})=M_0+M_1k_x^2+M_2k_y^2+M_3k_xk_y$, $A(\boldsymbol{k})=A_0+A_1k_x^2+A_2k_y^2+A_3k_xk_y$, and $B(\boldsymbol{k})=Bk_z$. In the specified basis, the inversion and time reversal operations take the following representations: $\mathcal{P}=\tau_z$, and $\mathcal{T}=\tau_0 K$ with $K$ the complex conjugation operation. One can easily verify that the model $H_0$ respects both symmetries. From this model, we find that a nodal loop would appear in the $k_z=0$ plane under the condition that $A_3^2-4A_1A_2<0$. The parameters fitted from the first-principles result are $M_0=-0.0363 $ eV, $M_1=0.3215$ eV\AA$^2$, $M_2=0.8593$ eV\AA$^2$, $M_3=0.7538$ eV\AA$^2$, $A_0=-0.0355$ eV, $A_1=1.5961$ eV\AA$^2$, $A_2=1.3446$ eV\AA$^2$, $A_3=1.7215$ eV\AA$^2$, and $B=1.0906$ eV\AA, which satisfy the condition above. The fitting result of the band structure is shown in Fig.~\ref{fig8}(a).

When SOC is included, there will be additional SOC terms in the Hamiltonian. Treating SOC as perturbations, we add to model $\mathcal{H}_0$ in Eq.~(\ref{kp1}) the leading order symmetry-allowed SOC terms, given by
\begin{equation}\label{kp2}
\begin{split}
\mathcal{H}_\text{SOC}(\boldsymbol{k})&=\lambda_0k_z\tau_x\sigma_z+k_x\tau_x(\lambda_1\sigma_x+\lambda_2\sigma_y)\\
&+k_y\tau_x(\lambda_3\sigma_x+\lambda_4\sigma_y),
\end{split}
\end{equation}
where the Pauli matrices $\sigma_i$ stand for the real spin. Then the full model is
$\mathcal{H}=\mathcal{H}_0\otimes\sigma_0+\mathcal{H}_\text{SOC}$. One easily checks that the SOC terms open a gap at the original nodal loop. For CuTeO$_3$, the SOC strength is small, so the gap is also negligible.
The parameters obtained from fitting the first-principles result are $\lambda_0=0.0007 $ eV\AA,$\lambda_1=0.0012 $ eV\AA,$\lambda_2=-0.0084 $ eV\AA,$\lambda_3=0.0023 $ eV\AA,  $\lambda_4=0.0117 $ eV\AA. The fitting result for the band structure with SOC is shown in Fig.~\ref{fig8}(b).

\begin{figure*}[tbh]
\includegraphics[width=15cm]{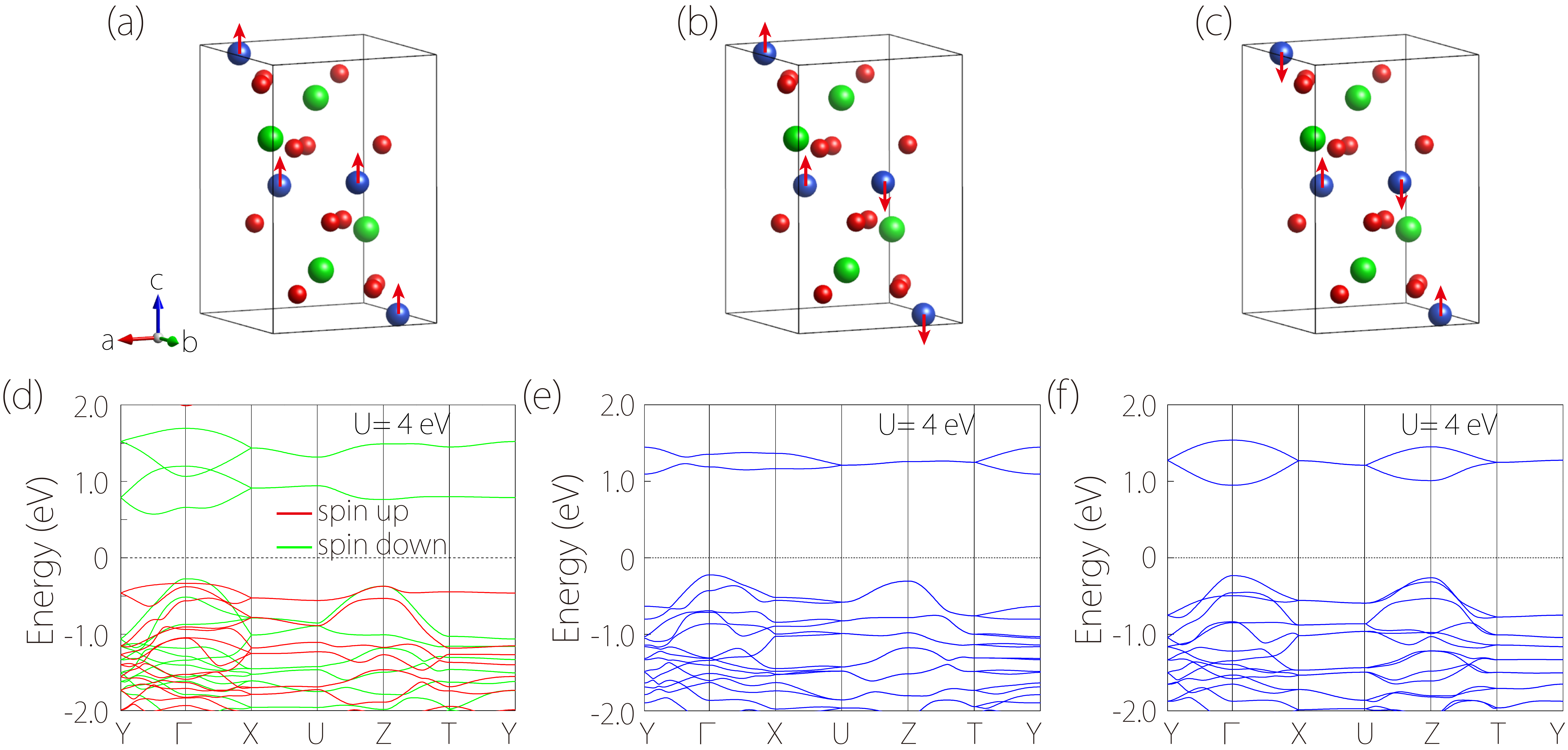}
\caption{Results from GGA$+U$ calculation. We consider three possible magnetic ground state configurations: (a) is for ferromagnetic, and (b,c) are for two antiferromagnetic configurations. (d-f) are the corresponding band structures for (a-c) with $U=4$ eV. SOC is not included here.}
\label{fig9}
\end{figure*}

\section{Discussion and Conclusion}

Based on their formation mechanisms, nodal loops can be divided into different classes. Certain nodal loops are accidental in the sense that their presence requires band inversion in certain regions of the BZ. And the loop can be adiabatically annihilated without breaking the system's symmetry. There also exist nodal loops that do not rely on the band inversion. In the absence of SOC, it has been shown that a 2D $\mathbb{Z}_2$ invariant can protect a nodal loop without band inversion~\cite{Fang2015,ahn2018}. Moreover, certain nonsymmorphic space group symmetries can guarantee the presence of nodal loops even in the presence of SOC~\cite{Chen2015a,Bzdusek2016,Wang2017,Li2018}. The nodal loop in CuTeO$_3$ here belongs to the first class, namely, it requires the band inversion around the $\Gamma$ point, and it can be annihilated when the band inversion is removed [as in Fig.~\ref{fig4}(a)] without changing the symmetry of the system.

In the band structure without SOC shown in Fig.~\ref{fig2}(a), one also notices that the bands are degenerate along the paths U-Z and Z-T, which lie on the $k_z=\pi$ plane. This behavior is not accidental. In fact, the bands on the whole $k_z=\pi$ plane must be doubly degenerate, forming a nodal surface in this plane. This is a Kramers-like degeneracy due to the anti-unitary symmetry $\mathcal{T}\widetilde{S}_{2z}$, where $\widetilde{S}_{2z}=\widetilde{\mathcal{M}}_{z}\mathcal{P}$ is a two-fold screw rotation along $z$. One checks that
\begin{equation}
(\mathcal{T}\widetilde{S}_{2z})^2=e^{-ik_z}.
\end{equation}
On the $k_z=\pi$ plane, we have $(\mathcal{T}\widetilde{S}_{2z})^2=-1$, which leads to the Kramers-like degeneracy and hence the nodal surface. This argument also shows that the nodal surface here has to be residing in the $k_z=\pi$ plane. Recently, nodal surface semimetals have been proposed in a few real materials~\cite{Zhong2016,Liang2016,wuweikang2017}. More detailed analysis regarding the nodal surface can be found in Ref.~\cite{wuweikang2017}.

When SOC is included, the nodal surface is split. (The nodal surface cannot exist in the current case because of the preserved inversion symmetry. See Ref.~\cite{wuweikang2017}.) However, four-fold degenerate Dirac points are observed at point Z (and also X), which are due to the presence of $\mathcal{P}$, $\widetilde{\mathcal{M}}_{z}$, and $\mathcal{T}$ symmetries at the point, analogous to those found in Refs.~\cite{Guan2017,Li2018}.

 We have assumed the paramagnetic phase for CuTeO$_3$ in this study. Since $d$ electrons in Cu may have correlation effects, we have also performed the GGA+$U$ calculations to investigate the possible magnetic phases. We find that for large enough $U$ value, a magnetic ground state is preferred. We consider a ferromagnetic (FM) configuration and two antiferromagnetic (AFM) configurations, as illustrated in Fig.~\ref{fig9}(a-c). For $U=4$ eV, we find that the AFM configuration in Fig.~\ref{fig9}(b) has the lowest energy among the three (each having $\sim 0.6 \mu_B$ moment per Cu site). In the band structures, all these magnetic states have a sizable bandgap larger than 1 eV. The FM state still has a nodal loop in the minority spin channel but it is above the Fermi level; whereas the loop is removed in the AFM states. Nevertheless, we note that: (i) Because the $3d$ electrons are less confined, the Hubbard $U$ correction often overestimates the tendency towards magnetism; (ii) No magnetism was observed in experiment at room temperature~\cite{pertlik1987}, and although there is no systematic experimental study of this material at lowered temperature, the Neel temperature for closely related compounds such as Cu$_3$TeO$_6$ ~\cite{li2017dirac} and Cu$_{3-x}$Zn$_x$TeO$_6$~\cite{bendaoud1994} are all below 70 K. Thus, our result presented here should be valid for the temperature range above the material's Neel temperature, which is not expected to be high.

In conclusion, based on first-principles calculations and symmetry analysis, we predict that the monoclinic CuTeO$_3$ is an almost ideal nodal-loop semimetal. There is a single nodal loop in the band structure close to the Fermi level. The loop is quite flat in energy, and there is no other extraneous band nearby. The drumhead-like surface states corresponding to the nodal surface are identified. We show that the loop is protected by two independent symmetries in the absence of SOC, and the loop can be effectively tuned or even annihilated by strain.
When SOC is considered, the nodal loop opens a tiny gap and the system (in a strict sense) becomes a $\mathbb{Z}_2$ topological metal with spin-polarized surface states. We have constructed a low-energy effective model to describe the nodal-loop phase and the effect of SOC. Since the SOC strength is very small, the monoclinic CuTeO$_3$ is well described as a nodal-loop semimetal. The bulk nodal loop and the surface states can be directly probed in the angle-resolved photoemission (ARPES) experiment. Our result offers a promising platform for exploring the intriguing physics associated with nodal-loop semimetals.

\begin{acknowledgements}
The authors thank X.-L. Sheng, S. Wu, and D. L. Deng for valuable discussions. {This work was supported by the National Key R$\&$D Program of China (Grant No.~2016YFA0300600), the MOST Project of China (Grant No.~2014CB920903), the NSF of China
(Grant No.~11734003 and 11574029)}, and the Singapore Ministry of Education AcRF Tier 2 (Grant No.~MOE2015-T2-2-144). Si Li and Ying Liu contributed equally to this work.
\end{acknowledgements}

\begin{appendix}

\section{Band structure result with modified Becke-Johnson potential}
\begin{figure}[t]
\includegraphics[width=8cm]{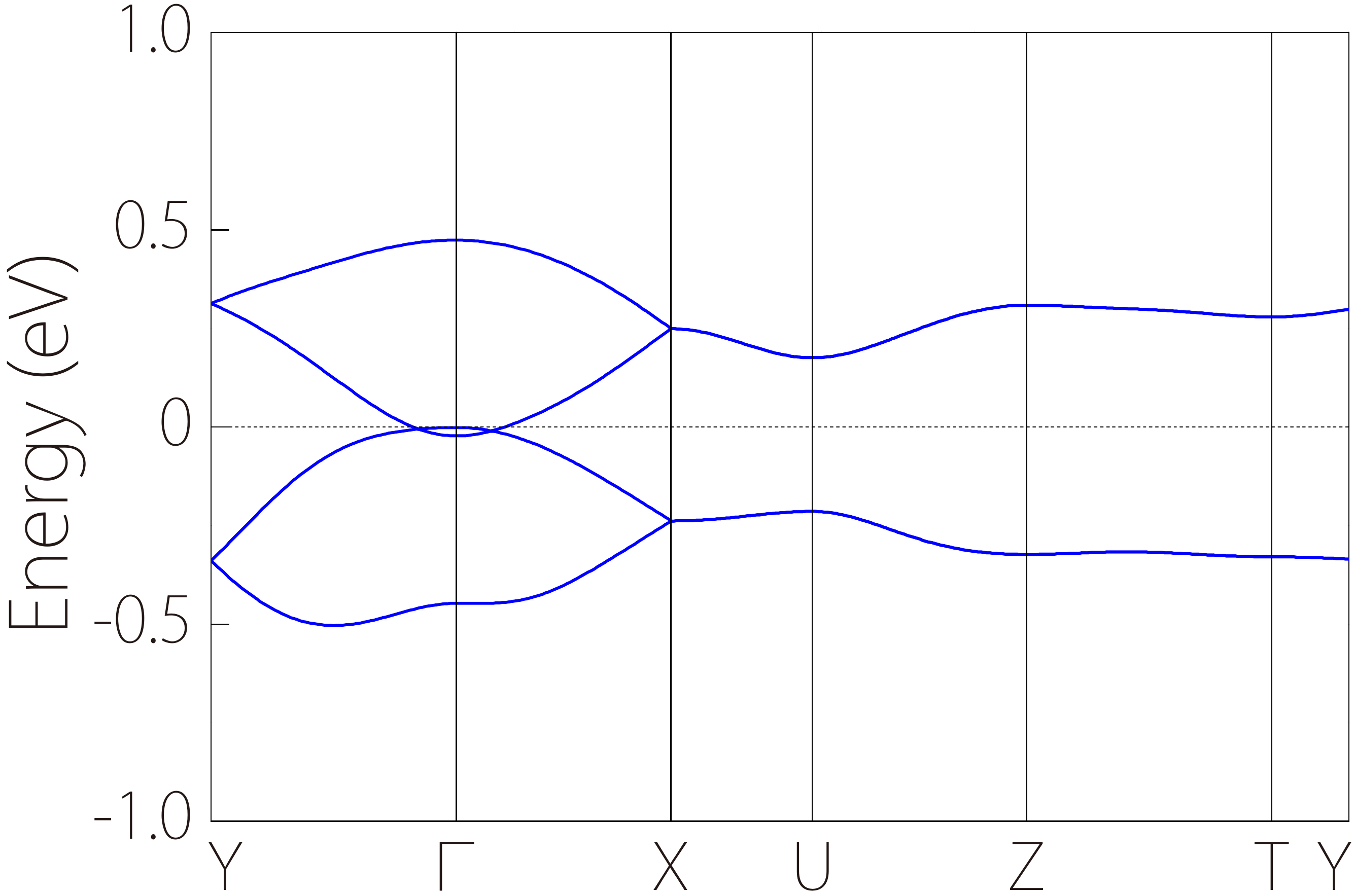}
\caption{Band structure of the monoclinic CuTeO$_3$ obtained using the mBJ approach (in the absence of SOC).}
\label{fig10}
\end{figure}

The band structure features are verified by using the more accurate approach with the modified Becke-Johnson (mBJ) potential~\cite{tran2009}. The result is plotted in Fig.~\ref{fig10}. The result indicates that the essential features including the nodal loop remain the same as the GGA result.

\end{appendix}


%

\end{document}